\documentclass[12pt,preprint]{aastex}

\usepackage{bm}
\usepackage{xcolor}
\usepackage{amsmath}
\colorlet{myc1}{green!20!red!80!}
\definecolor{rkka}{RGB}{219,66,32}

\newcommand{\eb}{\begin{equation}}
\newcommand{\ee}{\end{equation}}

\shorttitle{Functional principal component analysis of radio-optical reference frame tie}
\shortauthors{Makarov}
\begin{document}

\title{Functional principal component analysis of radio-optical reference frame tie} 
\author{Valeri V. Makarov}
\affil{United States Naval Observatory, 3450 Massachusetts Ave. NW, Washington, DC 20392-5420, USA}
\email{valeri.makarov@navy.mil}

\begin{abstract}
The Gaia optical reference frame is intrinsically undefined with respect to global orientation and spin,
so it needs to be anchored in the radio-based International Celestial Reference Frame (ICRF) 
to provide a referenced and quasi-inertial
celestial coordinate system. The link between the two fundamental frames is realized through two samples
of distant extragalactic sources, mostly AGNs and quasars, but only the smaller sample of radio-loud ICRF sources with
optical counterparts is available to determine the mutual orientation. The robustness of this link can be mathematically
formulated in the framework of functional principal component analysis using a set of vector spherical harmonics to
represent the differences in celestial positions of the common objects. The weakest eigenvectors are computed, which
describe the greatest deficiency of the link. The deficient or poorly determined terms are specific vector
fields on the sphere which carry the largest errors of absolute astrometry using Gaia in reference to the ICRF.
This analysis provides guidelines to the future development of the ICRF maximizing the accuracy of the link over the entire
celestial sphere. A measure of robustness of a least-squares solution, which can be applied to any linear
model fitting problem, is introduced to help discriminate between reference frame tie models of different degrees.

\end{abstract}


\section{Introduction}
\label{Introduction}
The latest releases of the Gaia mission data \citep[Gaia DR2, EDR3][]{pru,bro} are the currently best realizations of the
global optical celestial reference frame, which is a well-defined, quasi-inertial (non-rotating) system of celestial
coordinates serving as the basis for other subsidiary coordinate systems used in astronomy, celestial mechanics, and navigation.
In principle, the choice of coordinate triad is a matter of convention and convenience in practical
applications, as long as the coordinate system is well-defined and inertial, i.e., non-rotating with respect to distant
extragalactic sources. However, the existence of different reference frames dictates the establishment of certain
rigorous rules of comparison and data translation accounting for the existing misalignment.
The transfer of coordinates from one system to another is non-trivial. It is achieved either by direct position measurements
in reference to Gaia optical sources, or through a chain of intermediate observations that can establish the rules of this
transformation. A set of mathematical transformation rules is called a reference frame tie in this paper.
The reference system of the Gaia DR2 astrometric solution,
which defines the orientation of the coordinate triad in space and the rigid spin of the entire ensemble of 
1.7 billion objects, is not independent \citep{lin}. 
It was tied to the International Celestial Reference Frame (ICRF) as accurately as possible using a preliminary solution
for the version 3 \citep[preliminary ICRF3,][]{cha} \citep{mig18}. The spin, on the other hand, was adjusted to a nearly zero net value using a
much greater set of infrared-detected quasi-stellar objects (QSO) and active galactic nuclei (AGN), only a small
fraction of which are radio-loud sources observable with the VLBI. The prior information in this case is that the true proper motions
of extragalactic sources are expected to be unmeasurably small because of their great distances from us. There are physical
effects violating this assumption including the {\it secular aberration} \citep[e.g.,][]{kov, kop} at the level of
a few microarseconds per year, which cannot yet be measured with confidence. Rapid variations of the underlying
structure of radio-loud quasars can cause apparent motion of VLBI-observed cores and position differences
due to difference of observation epochs. Even greater randomly directed apparent shifts
affecting the RORF tie can be caused by
the frequency-dependent position of the core radio emission, as measured by the VLBI, along the parsec-scale radio
jets \citep{pla19}. As far as the transformation of celestial positions
is concerned, there is no viable alternative to the Radio-Optical Reference Frame (RORF) objects. These are
radio-loud, unresolved AGNs of supreme astrometric quality, amenable to VLBI position measurements at the microarcsecond level,
with an addition of a handful nearby radio stars, which are mostly active binaries.
The RORF objects must be sufficiently optically bright to be observed by Gaia. 

The main requirement to the radio-optical reference frame tie is that it should be ``accurate". In practical terms,
this means that ideal instruments and telescopes should measure nearly the same coordinates of ideal celestial
sources at different wavelengths minimizing the overhead from the transformation rules. The sources used to establish
these rules are not ideal, however \citep{mak}. 
Both Gaia and ICRF positions are burdened by random and systematic (or sky-correlated) errors. While the random component
can be statistically estimated from the dispersion of observed position differences, the systematic part is more ambiguous
and difficult to trace. A sky-correlated error in either catalog, whether of instrumental or stochastic origin, degrades the quality
of the reference frame tie in the most direct way. A rigid rotation of the coordinate triad is one of the most intuitive
kinds of error. If the two systems are misaligned, positions determined in reference to the grid of radio sources
are systematically different from positions  determined by differential astrometry of optical sources.
In view of the crucial importance of the tie, much care has been taken in aligning both Gaia DR1 and DR2 systems
and the ICRF \citep{mig16, mig18}. There is no clearly defined recipe even for this limited adjustment, partly because of
the complex nature of the RORF objects, while the result is sensitive to the sample selection, weighting of data, and
other technical decisions one has to make  in this process.

The first Gaia data releases confirmed that a significant fraction of RORF quasars have radio-optical position
differences much beyond statistical expectation levels \citep{pet, mak16, pet18, mak19}.
In some cases, a careful one-by-one analysis of available images from ground-based surveys revealed a possible
reason for the discrepant positions. It was most frequently the contribution of the optical host galaxy component, which
may be non-symmetric with respect to the AGN due to intrinsic irregularities such as prominent dust structures.
Extended substrate images may perturb Gaia optical positions, but are not visible to radio-interferometric observations
with long baselines. 
A smaller number of cases were attributed to physical or visual duplicity (e.g., gravitational microlensing) or
plain confusion with a random field source. The impact of host galaxies should become negligible for greater distances
because of the ``fundamental plane"
relation \citep{ham}, where the magnitude
difference in $V$ between the host galaxy and the nucleus is larger for luminous nuclei. The binned median GaiaDR2$-$ICRF3
position offset indeed quickly declines with redshift $z$ at redshifts below 1, but appears to  rise again at $z\simeq
1.3$, which is hard to explain as the impact of extended optical structures \citep{mak19}. Filtering perturbed and extended
objects, statistical outliers beyond $3\sigma$ (99\% confidence) leaves only 2119 sources to establish the GaiaDR2-ICRF3 tie.

In this paper, the cleaned set of RORF objects (which is called the precious or vetted
set throughout the paper) is used to study in detail the GaiaDR2$-$ICRF3 tie,
which goes beyond the general misalignment of the coordinate triads but also captures large-scale relative distortions.
The position differences of the cleaned and vetted sample can be analyzed using a finite set of vector spherical harmonics (VSH), 
which are vector-valued functions
on the unit sphere, as briefly described in \S\ref{vsh.sec}. This analysis goes beyond the 
three first-degree magnetic VSH (rigid rotation)
and interesting patterns emerge for both position and proper motion vector fields. Do these patterns carry any statistical significance?
Two statistical quantities are introduced in \S\ref{rob.sec} called robustness and confidence to evaluate the quality of VSH fits,
and a simple way to improve it is discussed in \S\ref{wei.sec}. The paradigm of functional principal component analysis (FPCA)
in application to the reference frame tie determination is presented in \S\ref{fpca.sec}, with discussion and conclusions in 
\S\ref{con.sec}. The mathematical technique of principal component analysis (PCA) on a basis of 
orthogonal functions evaluated at a grid of data points is widely used in various research fields
and disciplines, including optimized modeling of complex 3D shapes \citep{det}, and in astronomy, for reduction of
photometric noise in 2D data collected with bolometric arrays \citep{dow} or processing of 1D sparsely sampled light curves
\citep{he}. In this paper, the FPCA method is applied to astrometric catalogs data on the basis of VSH functions.

\section{Vector spherical harmonics as a functional basis for reference frames}
\label{vsh.sec}

A continuous tangential vector field on the unit sphere (i.e., a two-dimensional vector-valued function of spherical coordinates)
can be uniquely represented as a linear expansion (or infinite linear combination) in vector spherical harmonics (VSH), which
are themselves orthogonal vector fields on the unit sphere. The orthogonality is realized via the Euclidean inner product
in the space of spherical functions, which involves integration over the surface with proper weights. The difference between
two catalogs can be viewed as a vector field $\boldsymbol{d}$ discretized on a set of fixed points  
with coordinates $\alpha$, $\delta$ \citep{bro66}
\eb
\boldsymbol{d}=[\Delta\alpha\;\cos\delta,\;\Delta\delta],
\ee
where $\Delta\alpha$ and $\Delta\delta$ are the angular coordinate differences, which are the equatorial coordinates in
this example\footnote{This paradigm can be used for any system of celestial coordinates.}. The basis vectors are
unique for each point, because they are the tangential directions at the given point completing the local coordinate triad,
often called the east and north directions \citep[see, e.g.,][for detailed description of the model]{mig12}. The vector
field $\boldsymbol{d}$ can be expanded in an infinite series of orthogonal (or orthonormal with the proper choice of
normalization coefficients) vector-valued spherical harmonics
\eb
\boldsymbol{d}(\alpha,\delta)= \sum_{l=1}^\infty \sum_{m=-l}^{l} (e_{l}^{m}\boldsymbol{E}_{l}^{m}+h_{l}^{m}\boldsymbol{H}_{l}^{m}),
\label{vsh.eq}
\ee
where we separate the two types of harmonics, the poloidal, or electric, functions $\boldsymbol{E}(\alpha,\delta)$, and
curl, or magnetic, functions $\boldsymbol{H}(\alpha,\delta)$, which are described in more detail in \citep{mamu}. The
three first order ($l=1$) magnetic harmonics represent the general solid rotation between the two catalogs. The same
representation may be used for the discretized proper motion
differences $\boldsymbol{\mu}=[\Delta\mu_\alpha\;\cos\delta,\;\Delta\mu_\delta]$,
in which case the first-order magnetic harmonics represent the overall spin difference of the two proper motion systems.
It is fairly straightforward to compute these expansions for a given cross-matched set of objects with coordinates and
proper motions in the two catalogs, but there are a few technical and methodological questions to be addressed first.

No proper motions are given in ICRF3 for RORF objects, while Gaia DR2 includes proper motions for most of them. Assuming
that the true proper motion of these mostly distant extragalactic sources are zero, the Gaia determinations are
purely observational errors. If they are completely random and uncorrelated between the objects, the net effect is
reduced by the square root of the number of components. The Gaia CRF is then expected to be quasi-inertial to the
precision of the low-order VSH coefficients, most notably, the first three magnetic harmonics of the proper motion field.
This precision is obviously dependent on the individual formal errors and the number of objects. As explained in
\citep{mak12}, the sky-correlated component of the vector field, which is captured by an expansion like
(\ref{vsh.eq}), arises from both random and systematic errors
of observations, so it cannot be equated with the systematic difference. In fact, the random component of a VSH
expansion can be dominating. If the VSH functions were truly orthonormal, the random error expectation would be evenly distributed
among the coefficients $e_{l}^{m}$ and $h_{l}^{m}$. The actual set of functions evaluated
at a finite set of points is never orthogonal because
of the nonuniform distribution of objects on the sky and the distribution of observational error among the objects \citep{mami}.
This gives rise to an error overhead in the fit caused the nonzero covariances between the VSH coefficients.

\begin{figure}[htbp]
  \centering
  \includegraphics[angle=0,height=0.30\textwidth]{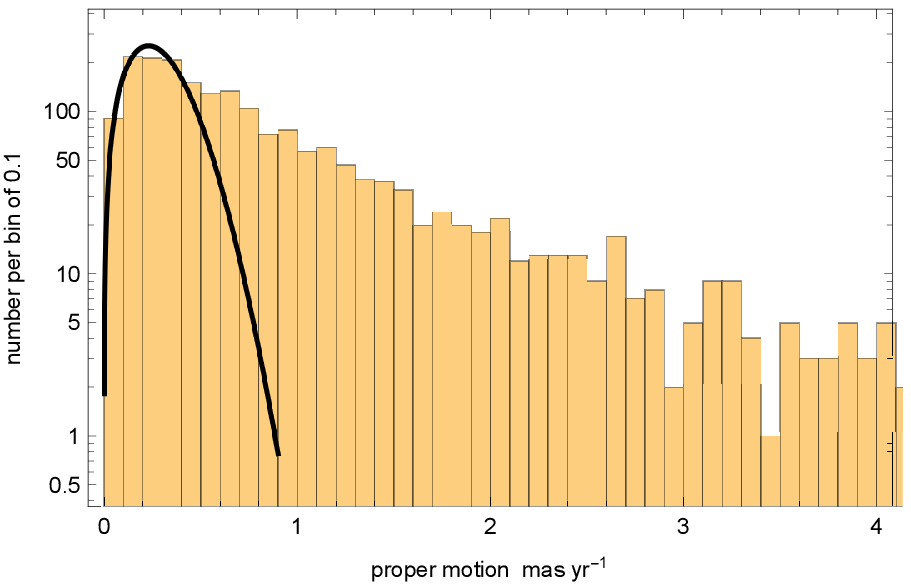}
  \includegraphics[angle=0,height=0.30\textwidth]{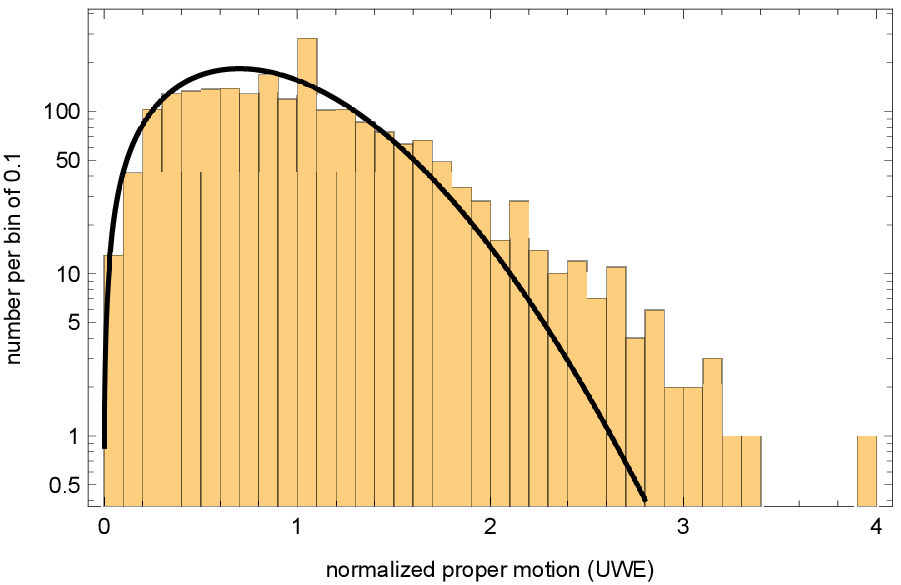}
\caption{Histograms of absolute (left) and normalized (right) proper motion magnitudes of the RORF sources from
the precious set. Note the logarithmic scale of both plots. The black lines show fitted Rayleigh distributions
with mean values 0.23 mas yr$^{-1}$ on the left and 0.7 on the right, only to assist the eye to distinguish
the core distribution and the tail of perturbed values. \label{pm_hist.fig}}
\end{figure}

How close is Gaia DR2 to the desired inertial reference frame in terms of the proper motion field? Fig. \ref{pm_hist.fig}
shows the histograms of absolute proper motion magnitudes in mas yr$^{-1}$ (left plot) and normalized proper motions
$\mu/\sigma_\mu$ (right plot) of the vetted ICRF3-Gaia RORF sources. 
The peaks of these distributions, where the majority of object reside,
can be fitted with Rayleigh distributions of $\nu=0.23$ mas yr$^{-1}$ and 0.7, respectively. Note that the latter number
for normalized differences is significantly smaller than the expected $\nu=1$. 
Either the formal errors of Gaia proper motions for these mostly
optically faint objects are overestimated, or, more likely, the adjustment of Gaia proper motions to the preliminary
version of ICRF3 took out a fraction of sky-correlated error. We would welcome this apparent reduction of dispersion,
except that it may also distort the real sky-correlated patterns in the motion of quasars, should they be present
at this level of precision. The powerful extension of the proper motion magnitude histogram beyond the Rayleigh core
simply reflects the distribution of Gaia formal errors for RORF sources, some of which are
marginally faint. There is also a small tail in
the sample distribution of normalized proper motions. This may be the manifestation of remaining astrometrically perturbed
objects in the precious set. We remind the reader, that the vetting procedure involved ICRF$-$Gaia position differences,
but was independent of proper motions.

Formally, if the VSH functions were orthogonal, and we could ignore the presence of outliers, the residual spin of DR2
with respect to ICRF3 would be $0.23\cdot \sqrt{3/n_{\rm obj}}=8.7$ $\mu$as yr$^{-1}$. This is how close we can get to
the inertial CRF paradigm with Gaia DR2 in the best case scenario. The number is too large to hope to detect the
effect of secular acceleration or signs of propagating primordial gravitational waves. However, a more reliable
estimation of uncertainties can be obtained from the full covariance matrices of position and proper motion VSH
coefficients.

\section{Robustness and confidence of VSH fits}
\label{rob.sec}
\textcolor{myc1}{}

Like numerous other problems of astrometry, and, generally, astronomy, finding a VSH fit to the observed vector field
$\boldsymbol{d}$ is an over-determined linear problem, which can be written as
\eb
\boldsymbol{A}\;\boldsymbol{x} = \boldsymbol{d},
\ee
where $\boldsymbol{A}$ is a design matrix, and $\boldsymbol{x}$ is the vector of unknowns, in our case, the coefficients
$e_{l}^{m}$ and $h_{l}^{m}$ in arbitrary order, which should be consistent with the order of columns in $\boldsymbol{A}$.
This system of equations is most commonly solved by the least-squares method, although other norms of the residual
vector could be considered, e.g., a 1-norm solution instead of the 2-norm. The regular 2-norm (least-squares) solution
to this problem is
\eb
\hat{\boldsymbol{x}}=(\boldsymbol{A}'\boldsymbol{A})^{-1}\boldsymbol{A}'\,\boldsymbol{d}.
\ee
The symmetric term of the pseudoinverse matrix $\boldsymbol{C}=(\boldsymbol{A}'\boldsymbol{A})^{-1}$ is called the covariance matrix of $\hat{\boldsymbol{x}}$,
which contains essential information about the statistical uncertainty of this solution. It is easily verified that
{\it if} the right-hand part of the system splits into a true vector $\bar{\boldsymbol{d}}$ and a random vector $\boldsymbol{\epsilon}$,
whose components are independent variables drawn from ${\cal N}(0,1)$, the expectation $E(\hat{\boldsymbol{x}}\,\hat{\boldsymbol{x}}')=
\boldsymbol{C}$. The diagonal elements of this matrix are traditionally given high importance, as they are {\it formal} variances
of the determined parameters. For example, all astrometric catalogs that are in use today list the square roots of the diagonal values
as standard deviation errors of the astrometric parameters, while starting with the Hipparcos catalog, the covariances have also
been published. Being a useful and intuitively understandable measure of solution quality, the covariance matrix has considerable
limitations. It represents the parameters of the statistical distribution of determined parameters reasonably accurately only
as long as the above assumptions hold. In practice, they are never realized, as the measurement errors are not normally
distributed, nor are they uncorrelated. The single measurement precision is not confidently known, but instead is roughly estimated
from shot noise statistics and assumptions about instrumental contributions. In many cases, the calculated covariance has to
be scaled by the normalized $\chi^2$ statistic from the post-fit residuals (this was implemented in the Gaia mission catalogs,
for example). This does not correct the solution for a suboptimal distribution of weights. Furthermore, as we will see in the following, a single realization of the residual vector may significantly deviate from the assumed mean statistics in cases of
correlated right-hand parts.

These issues come to the front when we attempt to describe the transformation between the radio and optical reference frames
in terms of VSH. Let us consider the singular value decomposition \citep[SVD,][]{gol} of the design matrix
\eb
\boldsymbol{A}=\boldsymbol{U}\;\boldsymbol{S}\;\boldsymbol{V}'.
\ee
The diagonal matrix $\boldsymbol{S}$ includes the singular values of $\boldsymbol{A}$ in decreasing order, $s_1\geq
s_2\geq \ldots \geq s_{m}$, with $m$ the total number of VSH terms in the fit. The least-squares solution is then
\eb
\hat{\boldsymbol{x}}=\boldsymbol{V}\boldsymbol{S'}^{-1}\boldsymbol{U}'\,\boldsymbol{d},
\ee
and the actual covariance of this solution is
\eb
{\rm Cov}(\hat{\boldsymbol{x}})=E((\hat{\boldsymbol{x}}-\bar{\boldsymbol{x}})(\hat{\boldsymbol{x}}-\bar{\boldsymbol{x}})')=
\boldsymbol{V}\boldsymbol{S'}^{-1}
\boldsymbol{U}'\;{\rm Cov}(\boldsymbol{d})\;\boldsymbol{U}\boldsymbol{S}^{-1}\boldsymbol{V}',
\ee
which is equivalent to the previous definition of covariance
\eb
\boldsymbol{C}=\boldsymbol{V}\;\boldsymbol{S}^{-2}\;\boldsymbol{V}'
\ee
{\it only if} ${\rm Cov}(\boldsymbol{d})$ is identity. Here the square $m\times m$ diagonal matrix
$\boldsymbol{S}^{-2}=(\boldsymbol{S}'\boldsymbol{S})^{-1}$ is composed of numbers $s_j^{-2}$.
We are also interested to know how far this solution may deviate
from the true parameter vector. We would like to minimize the 2-norm squared of this deviation,
\eb
r^2= (\hat{\boldsymbol{x}}-\bar{\boldsymbol{x}})'(\hat{\boldsymbol{x}}-\bar{\boldsymbol{x}})= 
(\boldsymbol{S}'^{-1}\boldsymbol{U}'\boldsymbol{\epsilon})'
(\boldsymbol{S}'^{-1}\boldsymbol{U}'\boldsymbol{\epsilon}),
\label{r.eq}
\ee
where we introduced the error of the right-hand part $\boldsymbol{\epsilon}=\boldsymbol{d}-\bar{\boldsymbol{d}}$. The
absolute error of the solution vector is the projection of the right-hand part error onto the basis of singular
vectors $\boldsymbol{u}_i$, which are the columns of $\boldsymbol{U}$, weighted with the reciprocal singular values
$s_i^{-1}$. These weights, which define the contribution of different components of the measurement error to the
resulting solution error, are key to understanding the properties of the fundamental reference frame tie.  

Eq. \ref{r.eq} is valid for any error vector $\boldsymbol{\epsilon}$, including a non-random, or systematic, component.
The absolute quadratic error of the frame tie is, generally,
\eb
r^2_{\rm arbitrary}=\sum_{j=1}^m s_j^{-2} (\boldsymbol{u}_j \cdot \boldsymbol{\epsilon})^2,
\label{rab.eq}
\ee
where $\boldsymbol{u}_j$ are the orthonormal column vectors of $\boldsymbol{U}$.
It is straightforward to prove that the largest absolute error is realized on the n-dimensional manifold of unit vectors $\boldsymbol{\epsilon}$
when it coincides with the least significant eigenvector, i.e., $\boldsymbol{\epsilon}=\boldsymbol{u}_m$. This is rather
improbable for a random data error, but there is a certain probability that $\boldsymbol{\epsilon}$ happens to be
rather well aligned with this specific eigenvector. In other words, the absolute error of a least-squares solution with
random noise in the data is a random variable itself, and a particular outcome depends on the specific realization of the measurement
error. 

For a purely random error $\boldsymbol{\epsilon}$ and optimally weighted least-squares problem, 
the mathematical expectation of $r^2$ is 
\eb
E[r^2_{\rm random}]=e_0^2\;\sum_{j=1}^m s_j^{-2}
\label{rrandom.eq}
\ee
with $e_0$ denoting the single measurement error of unit weight, i.e., ${\rm var}[\epsilon_i]=
e_0^2$, $i=1,\ldots, n$. With respect to random errors in the input data,
the accuracy of the frame tie is completely determined by the single measurement error and the spectrum of singular
values $s_j$. In practice, an additional overhead of solution error is caused by our imperfect knowledge of the
measurement variances, i.e., non-optimal weights in the least-squares adjustment.What we usually understand by solution precision or uncertainty of a solution is only the expectation of the average outcome
from a great number of trials. In practice, we are given only one trial in most cases, and the solution can be by chance 
much worse than the expectation
even when the usual characteristics, such as the trace of the covariance matrix, appear to be acceptable. Indeed,
the same value of trace
\eb
{\rm trace}[{\rm Cov}(\hat{\boldsymbol{x}})]=\sum_{j=1}^m s_j^{-2}
\ee
can be realized when all the singular values are equal, which is the best possible or optimal situation, or a fraction of
them are much smaller than the others, resulting in a finite probability of a catastrophically inaccurate solution. Therefore,
a metric of solution accuracy and reliability should include not only the range of singular values, which is reflected
in the traditional condition number $s_1/s_m$, but also the greatest singular value itself. The following dimensionless
parameter, called {\it robustness} of a least-squares solution, is found to represent well the desired uniformity
of the singular value spectrum and sensitivity to systematic perturbations in the input data:
\eb
{\cal R}=\frac{\sqrt{n}\,s_1^{-1}}{\sqrt{\sum_{j=1}^m s_j^{-2}}},
\label{rob.eq}
\ee
where $n$ is the number of condition equations.

This metric can be applied to any astronomical least-squares problems, and can be compared between different applications.
It captures the risk of bad solutions due to the spread of singular values, but also reflects the general precision of input data, 
as well as the number of conditions. For the best possible condition of a unitary matrix
$\boldsymbol{A}$, the robustness value becomes ${\cal R}=\sqrt n/\sqrt m$, which reflects how much the system is overdetermined.
This sets the upper bound of robustness, because the value can only be smaller for a non-unitary condition matrix.
To achieve a higher robustness, the possible actions are 1) obtain more measurements to increase $n$; 2) minimize the number
of unknowns $m$; 3) improve the structure of condition equations
by setting up more sensitive measurements (for example, take parallax measurements at the times of maximum parallactic
excursion) to minimize the spread of singular values. The robustness parameter captures both the compression of a random
component and the chances of particularly poor solutions in the presence of systematic errors.

When using VSH functions to establish the radio-optical frame transformation, we encounter the classic problem of model
fitting in astrometry. The series in Eq. \ref{vsh.eq} is infinite, but the model should be finite, so where do we
truncate the series? Would ``the more the better" principle work here? Both theory and practice of astrometric data
reductions prove this idea to be faulty. 

The residual vector $\boldsymbol{\rho}=\boldsymbol{d}-\hat{\boldsymbol{d}}$ is the product of the idempotent matrix 
$(\boldsymbol{I}-\boldsymbol{U}\,\boldsymbol{U}')$ and vector $\boldsymbol{d}$, so that the squared norm of the residual
is
\eb
\boldsymbol{\rho}'\,\boldsymbol{\rho}=\boldsymbol{d}'(\boldsymbol{I}-\boldsymbol{U}\,\boldsymbol{U}')\boldsymbol{d}.
\label{rho.eq}
\ee
The square idempotent matrix $\boldsymbol{U}\,\boldsymbol{U}'$ has a rank of $m$, hence, in the diagonalized form, it
is equivalent to $\boldsymbol{Q}\,\boldsymbol{D}\,\boldsymbol{Q}'$, with $\boldsymbol{Q}$ being an orthogonal matrix,
whose columns are the eigenvectors of the diagonalization mapping, and $\boldsymbol{D}$ a diagonal matrix with
exactly $m$ elements equal to 1, and the rest equal to 0. If $\boldsymbol{\epsilon}$, which is the error in $\boldsymbol{d}$,
is random and ${\rm Cov}(\boldsymbol{\epsilon})=\boldsymbol{I}$ (due to the optimal weighting of the condition equations),
we readily arrive for an unbiased LS solution at the expected squared norm of the post-fit residual vector 
\eb
E[\boldsymbol{\rho}'\,\boldsymbol{\rho}]=n-m.
\ee
Therefore, the RMS post-fit residual usually declines as $\sqrt{n-m}$ as we include more fitting functions in the model.
This behavior is a direct consequence of the independence of the residual norm (Eq.~\ref{rho.eq}) on the singular values $s_j$, which
specify the actual strength of the conditions, or the compression factors for the eigenvector components of a right-hand
side perturbation. By including additional fitting functions in the model, we may get adverse results (i.e., less accurate
solution) if these functions provide a weaker condition and result in generally smaller or more dispersed $s_j$. The
condition factor $\sum_{j=1}^m s_j^{-2}$, which emerges in the expected absolute error of the parameter vector, Eq. \ref{rrandom.eq},
is a viable alternative to post-fit residuals. The well-known statistical $F$-test also provides useful guidance about where
to truncate the fitting model. Based on the observation that the quadratic form in Eq. \ref{rho.eq} is $\chi^2$-distributed
due to the mapping matrix being idempotent, one can compare the significance of additional fitting functions in terms of the
confidence intervals:
\eb
{\rm Conf}[m_2,m_1]=F_{\rm CDF}\left[\frac{(\chi^2_{m_1}-\chi^2_{m_2})(n-m_2)}{(m_2-m_1)\chi^2_{m_2}}\right],
\label{conf.eq}
\ee
where $\chi^2_{m}$ is the squared norm of normalized post-fit residuals (Eq. \ref{rho.eq}) with $m$ fitting functions in
the model, $F_{\rm CDF}$ is the standard cumulative distribution function (CDF) for the $F$-distribution, $m_2>m_1$ are
the numbers of terms in the two models under comparison. An improvement in the post-fit residuals is deemed statistically
significant if thus calculated ${\rm Conf}[m_2,m_1]$ is greater that a certain limit ${\rm Conf}_{\rm lim}$. One of the
shortcomings of this estimate is that the choice of ${\rm Conf}_{\rm lim}$ is somewhat arbitrary, mostly reflecting the
subjective tolerance to model overfitting. In practice, the computed CDF often saturates in overdetermined problems,
losing the capability to discriminate between alternative models. The other shortcoming is that this statistic is
based on the observational residuals and has a weak relation to the accuracy of the solution vector.

\section{Functional PCA of the reference frame tie}
\label{fpca.sec}

The spectrum of singular values of the weighted design matrix defines the overall accuracy of a LS solution, as well
as the proposed robustness parameter ${\cal R}$ (Eq. \ref{rob.eq}). It can be explored in a deeper analysis of the
quality of the reference frame tie. We note that the projections of the data error vector onto the basis of singular
vectors $(\boldsymbol{u}_j \cdot \boldsymbol{\epsilon})$ in Eq. \ref{rab.eq}, which define the {\it absolute} error
of the solution, are weighted by the reciprocal singular values. Therefore, the greatest value $s_1$ identifies the
{\it most significant} singular vector $\boldsymbol{u}_1$, which can be most accurately determined from the given set
of data. Conversely, the least significant singular vector is $\boldsymbol{u}_m$, and the solution becomes very poor
if the data error vector is aligned with it. The singular vectors are conventionally called principal components (PC) of
a linear system. In most applications, a few most significant PCs are of interest, because they provide a sufficiently
accurate, but less noisy, representation of the observed data set with a given model. In the case under investigation,
the singular vectors are linear combinations of vector spherical harmonics, which are vector-valued continuous functions
on the unit sphere---thus, the PCs are in fact functions that are orthogonal on the specific set of data points\footnote{
In practice, adding or removing a single source from the list of radio-optical reference frame objects changes the
entire set of principal components}, and their
analysis becomes a functional principal component analysis (FPCA). 

If we are interested in astrophysical or cosmological applications of the fundamental reference frame, such as the
secular acceleration of the Solar system \citep{kov, kop, kli} or the presence of primordial gravitational waves, the 
functional PCA can be applied too. The signal is confined to a certain subset of VSH. The most
direct way is to solve the input vector field for that specific set and compute the covariance matrix. However,
the resulting uncertainties or the robustness value $\cal{R}$ (Eq. \ref{rob.eq}) would not account for the
potential weakness originating from those principal components of the RORF tie that have large coefficients $s_j^{-2}$
but are not explicitly present in the signal estimation. It may be advisable to compute the projections of the given
VSH set onto the RORF tie functional PCs and, if necessary, regularize the solution by accepting only the most 
significant components of this decomposition. The motivation is that any portion of the signal that is carried
by the poorly determined RORF tie PCs cannot be trusted to have an acceptable level of input error.

In the context of this paper, the least significant PCs should be identified, because they represent the weakest
part of the reference frame tie. Indeed, if $\boldsymbol{\epsilon}=\boldsymbol{u}_m$, the error propagates
with the factor $s_m^{-1}$ into the solution, which may result in substantial error magnification. 
This dangerous situation becomes especially probable when the model is badly overfitted, i.e., the number
of VSH terms is unreasonably high.  

\subsection{Numerical experiments with Gaia DR2 and ICRF3 data}

We start with the ``precious set" of 2119 RORF objects selected from the overlap of the ICFR3 and Gaia DR2 catalogs
\citep{mak19}. This sample includes only those radio-loud sources from the S/X catalog that have optical counterparts
in Gaia DR2 within 3 radii of the combined error ellipse, which corresponds to the 99\% confidence level. The selection
process also relied on a careful vetting of available high-quality images from the Pan-STARRS and Dark Energy Survey (DES) archives.
In view of the unexpectedly large fraction of astrometrically perturbed sources (approaching 1/4), only the best
quality data should be used to establish the RORF tie. The results may be crucially sensitive to statistical
outliers, as we will see in the following.

A series of LS solutions was produced for the position differences DR2$-$ICRF3 with VSH degrees $L$ between 1 and 8.
The number of fitting VSH functions is $m=2(L+1)^2-2$. Each object generates two condition equations because of the
two-dimensional character of the data. Apart from the position difference fit, a similar solution with the same
VSH degree is obtained for Gaia proper motions\footnote{Only 1954 RORF objects in the precious set
have Gaia proper motions}, which are expected to be zero. For each of these trials, the robustness (Eq. \ref{rob.eq})
and confidence (Eq. \ref{conf.eq}) are computed, the latter for $m_1=0$, $m_2=2(L+1)^2-2$, thus estimating the
significance of the post-fit reduction in residuals to the input data. The derived confidence levels can then be understood as
the probability of rejecting the null hypothesis (that the two reference frames are statistically identical).
The magnitude of the fitted vector field is
characterized with the median and maximum lengths of the solution vectors on the sample of RORF sources.

The results for the robustness and confidence parameters are presented in Table 1. For the unweighted position
differences, the robustness of VSH fits rapidly declines at low degrees, but this behavior slows down by $L=7$
(or $m=126$), where it becomes marginally acceptable. This is to be expected, because as the number of fitting VSH
terms increases, the singular values become smaller in general, and the high-degree harmonics cannot be accurately
determined from the available data. From this point of view, one would prefer to stay with the lowest possible
degrees. The estimated confidence, on the other hand, is not monotonic with $L$, and the most confident fit is
obtained at $L=3$. It still falls short of the normally accepted level of 0.99 corresponding to a confident fit.
The reason is the presence of a small number of large outliers, which perturb the solution, but continue to show
large post-fit residuals. Confident RORF tie solutions cannot be obtained with the original unweighted data
even after all the careful filtering and vetting performed in the construction of the precious set.

\begin{deluxetable}{lrrrrrr}
\tablecaption{RORF tie solutions for Gaia DR2 and ICRF3 \label{tie.tab}}
\tablewidth{0pt}
\tablehead{
\multicolumn{1}{c}{$L$}  &
\multicolumn{2}{c}{unweighted positions}  &
\multicolumn{2}{c}{weighted positions}  &
\multicolumn{2}{c}{weighted proper motions}  
  \\
\multicolumn{1}{c}{ }  &
\multicolumn{1}{c}{Robustness}  & \multicolumn{1}{c}{Confidence}  &
\multicolumn{1}{c}{Robustness}  & \multicolumn{1}{c}{Confidence}  &
\multicolumn{1}{c}{Robustness}  & \multicolumn{1}{c}{Confidence}  
  \\
}

%\rotate 
\tabletypesize{\scriptsize} \startdata
1 & 24.046 & 0.853 & 23.450 & 1.00 & 22.176 & 0.760\\
2 & 13.461 & 0.800 & 12.808 & 1.00 & 11.967 & 0.305\\
3 & 9.306 & 0.976 & 8.778 & 1.00 & 8.216 & 0.264\\
4 & 7.112 & 0.958 & 6.602 & 1.00 & 6.237 & 0.016\\
5 & 5.276 & 0.928 & 5.276 & 1.00 & 4.996 & 0.003\\
6 & 4.655 & 0.952 & 4.241 & 1.00 & 4.095 & 0.003\\
7 & 3.919 & 0.893 & 3.504 & 1.00 & 3.435 & 0.001\\
8 & 3.368 & 0.674 & 3.002 & 1.00 & 2.944 & 0.001\\
\enddata
%\tablenotetext{\dag}{Columns:\newline}
\end{deluxetable}

\subsection{Empirical weighting schema}
\label{wei.sec}
Insufficient confidence makes the RORF tie solution ambiguous because it may imply that the two celestial frames are
identical within the error of observation, and one could just as well ignore the VSH fits. There are reasons to
believe that this is more likely an artifact due to flukes in the data rather than a desired equivalence of the two
frames. Although a preliminary version of the ICRF3 catalog was used to align the Gaia DR2 CRF, this adjustment only
included the rigid rotation of the Cartesian coordinates, i.e., the three $L=1$ magnetic harmonics. We still find
significantly nonzero coefficients for these model terms, which shows how sensitive the result may be to the
selection of RORF objects and the weighting principle. The traditional approach of using the formal position uncertainties
and covariances (error ellipses) for weights is not well justified in this case. This may only capture the larger
expected errors due to faintness of some quasars in the precious set, or a smaller number of available astrometric
observations because of the non-uniformity of the scanning schema, but we know that the input positions are commonly
perturbed by some kind of ``cosmic error", which is not correlated with these technical circumstances. In establishing
a RORF tie, one would like to reduce the impact of this yet unknown error as much as possible. Using the precious set
with limited normalized offsets is the first step in this direction. Further improvement can be obtained with a
conservative empirical weighting technique based on the tie fitting residuals.

The unweighted VSH solution at a given $L$ is used to compute post-fit residuals, which constitute a discretized
vector field on the unit sphere. The magnitude of these vectors depends on $L$ in general, or the rank of the
design matrix, as discussed in \S~\ref{rob.sec}. The median residual arrow in position, for example, increases from
486 $\mu$as at $L=1$ to 533 $\mu$as at $L=4$, and further to 593 $\mu$as at $L=8$. This happens because a small
fraction of objects with discordant and perturbed position differences in the two catalogs pull the
unweighted VSH fit and make the rest deviate more from the solution. The median residual vector
length $p_{\rm med}$ is computed without weights, and in the next iteration, the weight is defined for $i$th object as
\begin{align}
w_i&=  1, & \: {\rm if} \:p_i\leq p_{\rm med} \nonumber \\
   &= 
   p_{\rm med}/p_i, & \: {\rm if} \:p_i> p_{\rm med} \nonumber 
\end{align}
The procedure can be iterated a few times, but numerical experiments showed that the convergence is very fast,
and two iterations are usually sufficient. Effectively, this schema down-weights the data points that are
less compliant with the smooth tie field, judging them less reliable for a reason that is not a priori known.

The results obtained with this weighting of data are listed in Table~1. The confidence of the tie solutions remarkably
improved across the range of VSH degrees, saturating at 1. Hence, the RORF tie leads to a statistically significant
reduction of the overall position misalignment between the two reference frames with this weighting schema.  Unfortunately,
this parameter does not tell us which set of VSH functions represents the tie most accurately. The robustness value
still drops with increasing $L$, and it is generally lower than that for the unweighted solution. This interesting
outcome may be interpreted  in different ways, with one possibility being a degradation of condition through the
least significant PCs. As will become clear in the next Section, the weakest condition comes from the distribution
of objects on the sky, caused by the irregular coverage of VLBI measurements. The additional perturbation of the
RORF objects that contribute the most to the determination of the least significant PCs may be coming from the
radio-interferometric side of the input data.

The last two columns in Table 1 give the estimated parameters  for the Gaia DR2 proper motion field, for the sake
of completeness, although proper motions are not part of the RORF tie. The robustness is slightly smaller for the
proper motion fits, which is caused by the smaller number of data points. The most striking result is that
the confidence is unacceptable even at the smallest number of fitting terms. In other words, the VSH fit
cannot be trusted in any case, because it does not provide the expected reduction in post-fit residuals.
This result confirms that the Gaia DR2 proper motion field for ICRF3 objects does not contain any determinable signal of interest,
such as the secular aberration or gravitational waves.

\subsection{The least significant principal component}
\label{lpc.sec}

As we determined previously, the absolute error of a general LS solution, and of a RORF tie in particular, is defined
by the projections of a specific perturbation vector in the input data onto the basis of singular vectors. A catastrophically
poor solution can be realized if the perturbation happens to be close to the singular vectors at the end of the
range defined by the rank. The least-significant singular vector $\boldsymbol{u}_m$ and the corresponding singular
value $s_m$ are of special interest, therefore, as they define the weakest part of the solution. The smallest singular
value decreases with $L$, so that VSH fits with many terms become increasingly risky. This behavior is caused by
the presence of irregularities and holes in the sample sky coverage. Indeed, high-degree VSH have many small wiggles
on the sphere corresponding to variations of higher spatial frequency. A dearth  of data at specific locations makes 
some of the high-degree model terms poorly conditioned. This also accounts for the rapid decrease of robustness $\cal{R}$ with $L$.

\begin{figure}[htbp]
  \centering
  \plotone{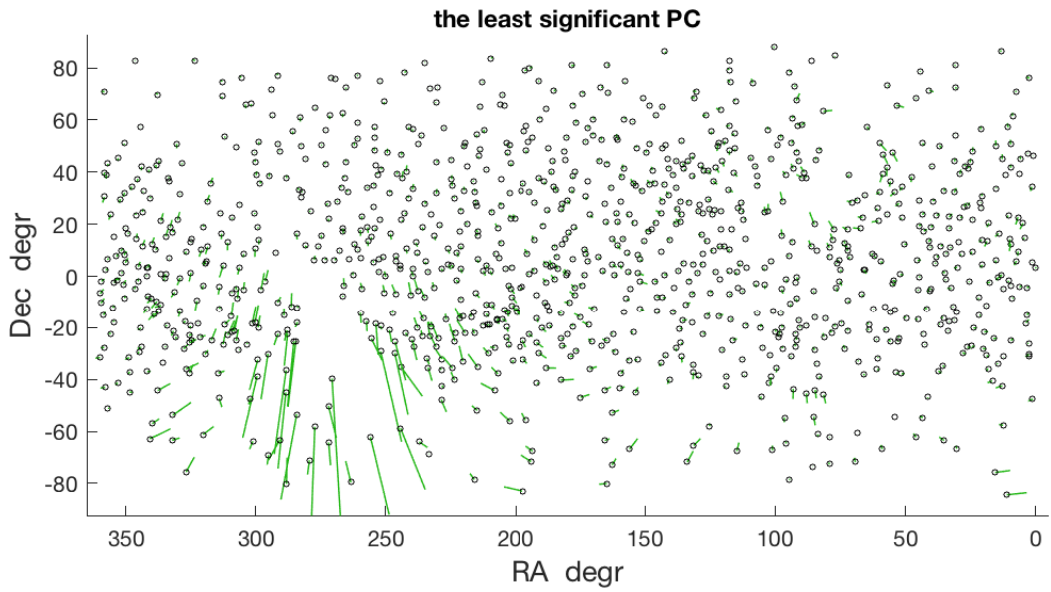}
\caption{Sky map of the least significant principal component of the RORF tie at degree $L=7$in equatorial coordinates. 
Small squares mark the position of radio-loud sources from the ICRF3 S/X catalog and their Gaia DR2
counterparts used to establish the tie. The arrows are arbitrarily scaled in length, representing a singular
vector field of unit length. \label{pc.fig}}
\end{figure}

Fig. \ref{pc.fig} gives a graphical presentation of the RORF tie least significant principal component (with empirical
weighting as described in \S~\ref{wei.sec}) at $L=7$. The robustness of this solution is just above 3.5 (Table 1),
so it is deemed the highest degree fit that is still acceptable. The arrows show arbitrarily scaled values of
$\boldsymbol{u}_{126}$ for half of the sample, to avoid overcrowding. The small circles represent the source positions,
i.e., the arrows' origins. The corresponding singular value is $s_{126}=15.2$, which indicates a nice degree of
measurement error shrinkage in the VSH fit even for this component. The graph reveals that the weakest component
of the Gaia-ICRF3 tie is confined to a specific area in the southern hemisphere, where the density of ICRF3 is
much lower, and a fairly large hole around the Galactic center.

\section{Conclusions}
\label{con.sec}
A wide range of basic research projects in astronomy, cosmology, and physics, as well as practical application in
geodesy and navigation, require position and motion determinations in the two fundamental reference frames. The consistency
of these frames is absolutely essential, and the transformation of coordinates between the frames, called the radio-optical
reference frame tie, should be established and registered as accurately as possible. On the example of the Gaia DR2 and ICRF3
S/X catalogs, it is shown in this paper, that a confident solution can be obtained after a careful two-stage vetting of
the available RORF sources. The first step is to minimize the impact of the cosmic error, which affects at least 25\% of
the sample, and may be caused by the internal structure of beamed radio jets \citep{pet18}, optical offsets of AGNs from the host galaxy's
centers, morphology of nearby galaxies, unresolved duplicity, gravitational lensing, and other phenomena. The second step
is to empirically downweight the remaining perturbations, which distort the functional fit, generate larger median post-fit
residuals, and drive the confidence of global solutions to below the acceptable levels. 

The newly introduced robustness parameter and the $F$-test confidence are helpful in determining the highest degree of
the VSH model, which provides the finest detail of the functional representation, while remaining reasonably accurate
and robust. It is demonstrated that the degree should not be larger than $L=7$ for the two investigated catalogs. The
choice of a lower degree is a trade-off between the desired robustness of the tie and the small scale fidelity in
applications limited to specific regions of the celestial sphere. In the context of general alignment
of the two fundamental celestial reference frames, when the best result at various spatial scales is desired, the recommended
degree is 6 or 7 for ICRF3 and Gaia DR2. This number may change upwards for the future Gaia, and, eventually, ICRF
versions as the quality of astrometric data improves.

The least significant functional principal component of the RORF tie demonstrates the weakest point, through which
the largest possible error can be realized. It is a specific vector field on the celestial sphere, and its largest vectors
are limited to well-defined areas of the sky, mostly situated in the southern hemisphere close to the Galactic plane
and the Galactic center. Clearly, the weakness is caused by the dearth of ICRF3 sources in the south, and the smaller
density of optical counterparts in Gaia due to crowding and extinction. The situation can be improved by prioritizing
the planned extension of ICRF defining sample in the southern hemisphere according to the distribution of the least
significant PC. The Gaia processing consortium may also consider making a special effort of windowing known ICRF
counterparts closer to the Galactic plane.

The Gaia DR2 proper motion field of the selected ICRF3 sources, on the other hand, does not contain any statistically
significant signal, so no transformation for the spin is required. This is to be expected, to some extent, because
the Gaia proper motion rigid rotation was adjusted on a much greater sample of infrared-detected quasars and AGNs \citep{mig16}.
The absence of significant higher-degree terms testifies to the good quality of astrometric calibration of time-dependent
parameters; on the other hand, the modest accuracy of the proper motion VSH fit precludes astrophysical exploration
of these data.

\section{Acknowledgments}
The author is grateful to Dr. Brian Luzum and the anonymous referee, whose comments and corrections
helped to improve the paper.
This work has made use of data from the European Space Agency (ESA)
mission {\it Gaia} (\url{http://www.cosmos.esa.int/gaia}), processed by
the {\it Gaia} Data Processing and Analysis Consortium (DPAC,\newline
\url{http://www.cosmos.esa.int/web/gaia/dpac/consortium}). Funding
for the DPAC has been provided by national institutions, in particular
the institutions participating in the {\it Gaia} Multilateral Agreement.

\end{document}